**A method for crystallographic mapping of an alpha-beta titanium alloy with nanometre resolution using scanning precession electron diffraction and open-source software libraries**

Ian MacLaren[1], Enrique Frutos-Myro[1,2], Steven Zeltmann[3,4], Colin Ophus[3]

1. School of Physics and Astronomy, University of Glasgow, Glasgow G12 8QQ, UK
2. School of Engineering, University of Glasgow, Glasgow G12 8QQ, UK
3. National Center for Electron Microscopy, Molecular Foundry, Lawrence Berkeley National Laboratory, 1 Cyclotron Road, Berkeley, CA 94720, USA
4. Platform for the Accelerated Realization, Analysis, and Discovery of Interface Materials (PARADIM), Cornell University, Ithaca, NY 14853, USA

**Abstract**

An approach for the crystallographic mapping of two-phase alloys on the nanoscale using a combination of scanned precession electron diffraction and open-source python libraries is introduced in this paper. This method is demonstrated using the example of a two-phase $\alpha$ / $\beta$ titanium alloy. The data was recorded using a direct electron detector to collect the patterns, and recently developed algorithms to perform automated indexing and analyse the crystallography from the results. Very high-quality mapping is achieved at a 3nm step size. The results show the expected Burgers orientation relationships between the $\alpha$ laths and $\beta$ matrix, as well as the expected misorientations between $\alpha$ laths. A minor issue was found that one area was affected by 180° ambiguities in indexing occur due to this area being aligned too close to a zone axis of the $\alpha$ with 2-fold projection symmetry (not present in 3D) in the Zero Order Laue Zone, and this should be avoided in data acquisition in the future. Nevertheless, this study demonstrates a good workflow for the analysis of nanocrystalline two- or multi-phase materials, which will be of widespread use in analysing two-phase titanium and other systems and how they evolve as a function of thermomechanical treatments.

## Introduction

Since its introduction in the 1990s[1, 2], EBSD has been hugely important in providing a method for crystallographic orientation mapping of materials. For example, in recent years crystallographic orientations in titanium alloys have been studied using electron backscatter diffraction (EBSD) in the SEM, which is very good at picking up the crystal orientation of the different $\alpha$ (hexagonal) and $\beta$ (body-centred cubic) areas in these complex microstructures and allowing detailed study of microscale or submicron crystallography as a result of various processes. Moreover, this has been used to study laths of $\alpha$ produced by precipitation from $\beta$ in different two-phase alloys (especially Ti-6Al-4V, but also Ti-6426 and others) and produced under different sequences of thermomechanical treatments. A particular area of interest has been what selection of variants is produced after any particular treatment[3-9].

However, EBSD does suffer from a lack of spatial resolution, whereby the resolution in the lateral direction may well be rather better (just a few nm) than the inclined direction (with the sample usually tilted at ~55° to the beam)[10, 11]. As such, reliably resolving finer features of a few tens of nm in size is difficult and orientation-dependent. Such nanostructures are, however, well within the range where scanned electron nanodiffraction can produce distinct diffraction patterns with a resolution down to a small number of nanometres (the exact spot size depends on probe convergence angle via the Abbe criterion, as normal). In recent years, scanned electron nanodiffraction, especially with the addition of precession (to form scanning precession electron diffraction, SPED) has been widely used for crystal orientation mapping[12], usually using some form of template matching algorithm. In this case, a large databank of possible diffraction patterns for the structure of interest are calculated to cover all crystallographically distinct orientations (with a spacing in orientation space to be determined by the user, using a trade-off between accuracy and calculation time / memory requirements). The experimental patterns are then compared with this databank one-by-one and the best correlation recorded in each case, which are then turned into a map of orientations. This has recently been used in the analysis of some of the nanoscale detail of twinning and deformation in Ti-Mo alloys[13, 14] and in pure Ti[15]. The downside of such work is that it has generally been performed using GUI-driven Windows software with no visibility of the source code to the users.

In recent years, the related techniques of scanned diffraction (perhaps more focussed on the diffraction patterns, and concerning well-separated diffraction spots from low convergence angle beams) and 4DSTEM (perhaps more focused on images calculated from the diffraction data, from higher convergence angle beams with overlapping diffraction disks) have had a huge boom in application[16]. The simple reason for this is the introduction of fast detectors, especially direct electron detectors, into electron microscopes and their integration into the acquisition control systems, allowing easy and fast acquisition of 4D datasets of a diffraction pattern at every point on a scan of a reasonable number of pixels [17]. In the field of orientation mapping, this has been successfully used with scanning electron nanodiffraction, and has been shown to give noticeable advantages over older indirect detectors in SPED[18]. Alongside this, there has been a significant growth in production of Open Source software for working with 4DSTEM and electron microscopy data[17], with a major advantage that all operations are transparent and can be examined in the source code. Also, these codes are optimised for use with the cleaner raw data coming from more recent detectors (not an image-processed export file with altered contrast). And when run in a notebook format, the notebook can be archived and made accessible to readers of a published study showing an exact documentation of what steps were applied in processing the dataset, the exact parameter choices and so on (something that is not often done when using GUI driven software). Moreover, as opposed to running GUI-driven software, it is far easier to be consistent in treatment of different datasets when using notebook or command-line driven operations from well-documented Open-Source software.

One such python library (*pyxem*) was recently shown to be useful for such orientation mapping on scanning electron nanodiffraction data on single phase datasets[19]. Additionally, another python library was recently introduced for working with 4DSTEM and nanodiffraction data, named *py4DSTEM* [20],. In the case of nanodiffraction, it can reduce spot diffraction patterns to lists of diffraction spot positions (so-called "points lists") and then perform calculations with these (such as strain). This has recently been applied to crystal orientation mapping with encouraging results so far, although the initial examples were on materials containing just one single crystalline phase[21]. In this paper, we demonstrate the use of this tool on a two-phase titanium alloy, which is an ideal test case for nanoscale mapping of a relatively complex structure. It is shown that this produces robust and reliable orientation mapping results. Additionally, these results are then analysed crystallographically in a quantitative manner using another Open Source python library, *orix*[22], and shown to match the expectations of crystallographic theory.

## Methods

A TIMETAL 550 (3-5% Al, 3-5% Mo, 1.5-2.5% Sn as main alloying elements, balance Ti) sample was prepared for TEM observation using a standard FIB liftout process using a FEI Nova Nanolab 200. Scanning precession electron diffraction was performed using a JEOL ARM200F scanning transmission electron microscope, equipped with a Nanomegas Astar scanning precession electron diffraction system controlled using TopSpin software. The smallest possible spot size and smallest condenser aperture (10 µm) were used to bring the probe current low enough to allow unsaturated data to be collected. Data was recorded using a Quantum Detectors MerlinEM direct electron detector, as detailed and used in our previous publications[18, 23]. A precession angle of 0.5° was used and the probe size was estimated at a diameter of 2.5-3 nm based on previous measurements for this condenser aperture; a 3 nm step size was used in data acquisition. The raw data was exported from the *.app5* file format used by TopSpin to an *hdf5* file using a suitable function in the *fpd* python library[24] (https://fpdpy.gitlab.io/fpd/). Automated crystallographic orientation mapping was performed using the *py4dstem* python library [20, 21], https://github.com/py4dstem/py4DSTEM, specifically using version 0.13.10, with the default correlation weighting scheme. Mapping was performed of the crystal orientations as two separate scans: firstly fitting to the $\alpha$ phase, and then to the $\beta$ phase, using the following cif files from the data of [25] for $\alpha$-Ti and [26] for V-stabilised $\beta$-Ti (.cif files from the Inorganic Crystal Structure Database, as provided by the Physical Science Data-science Service). Each fit was then exported to a .ang file of Euler angles for each scan position for each phase. The two *.ang* files were compared with a simple python script that determines the highest correlation index for each scan position and then writes a new .ang file with both phases included, with the phase with the highest correlation index chosen for each scan position (provided in the open data deposit associated with this paper). Analysis of the crystallographic orientation data in the .ang files was performed with the *orix* python library (version 0.10.2), https://orix.readthedocs.io/en/stable/. Cluster analysis was performed with the *DBSCAN* function of *scikit*-learn, with the maximum distance of two samples set to 15° (converted to radians), although the actual clusters were much more tightly packed in their orientations than this and were well separated and distinct in orientation from one another. Raw data and python notebooks are provided in an Open Data deposit associated with this paper showing the exact parameter choices made in each step of the analysis.

## Results

Figure 1 shows an overall view of the sample, and the specific area scanned using SPED. Figure 1a is a synthetic ADF survey image of the sample with the scan area for the higher resolution scan shown on there by a rectangle. The TIMETAL-550 material consists mainly of primary $\alpha$ with veins of residual $\beta$ between, which itself then contains some rather fine $\alpha$ laths (the Nb content of which was also mapped chemically in our previous work [27]) and the chosen scan area covered one of these $\beta$ veins with $\alpha$ laths. Figure 1b then shows a synthetic ADF image at higher resolution (pixel size of 3 nm) of this scan area, calculated using the integration of the intensity in the annulus shown overlaid on the average diffraction pattern (of the whole scan) shown in Figure 1c. This clearly shows laths which are a little darker against a slightly brighter background, which all makes sense, as the $\alpha$ laths are rich in Ti and Al (relatively light elements) and the β between them is enriched in rather higher-*Z* Nb, which will scatter more electrons to higher angles. (It should be noted that this is not to be compared to true HAADF contrast[28], which requires much higher inner angles for the annulus, but even relatively low angle annular dark field shows some Z-dependence, in addition to diffraction effects[17]). The laths are arranged in a number of orientations to the beam, but the diffraction pattern itself seems to be close to a $\langle 111 \rangle_\beta$ direction. Figure 1d is a roughly hexagonal diffraction pattern which would fit expectations for $\langle 111 \rangle_\beta$. Figure 1e shows the appearance of additional diffraction spots in one direction in a manner characteristic of a $\langle 2\overline{11}0 \rangle_\alpha$ direction, where we see $000n_\alpha$ and $hkin$ spots, where $n$ is odd, many of which only appear due to double diffraction (although kinematically forbidden), and this pattern can only be from α-Ti. Figure 1f is also hexagonal but is a little distorted from a regular hexagon in a way that turns out to match α-Ti better than β-Ti. This matching is shown in figure 2.

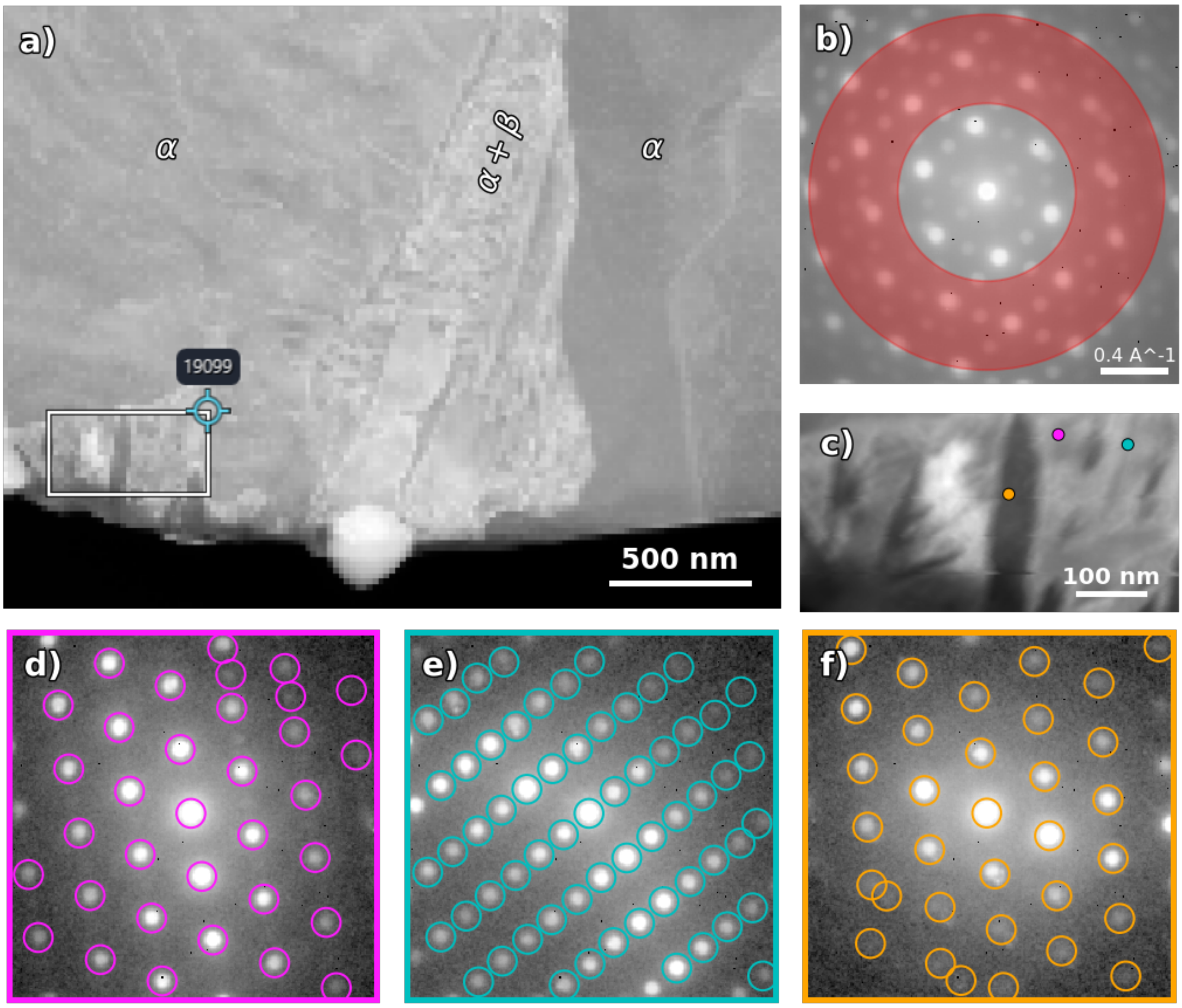

Figure 1: Overview of the area mapped with SPED and the basic data arising from this: a) General large area ADF view of an area containing mostly primary α grains with a thin strand of mixed α and β running through the middle and another α and β area in lower left that was mapped – the mapping box is shown; b) average diffraction pattern from the whole SPED dataset with the area used for calculating an ADF image highlighted in red; c) calculated virtual ADF image with three colour-coded points shown from whence individual diffraction patterns were extracted for display and diffraction spot detection; d) diffraction patterns with detected diffraction spots circled; d) was subsequently indexed as β and e) and f) were subsequently indexed as α.

Figure 2 shows an example of indexing for the points list derived from the diffraction pattern shown in Figure 1f, showing an overlay of the best fit pattern and the actual detected disks. Figure 2a shows the fit for α-Ti and Figure 2b shows the fit for β-Ti , both with the correlation indices printed. In this case, the correlation index is significantly better for α-Ti: although both phases can produce patterns that are qualitatively similar, the best fit β-Ti one is still noticeably distorted from the actual observed diffraction spot positions. Thus, it is clear that choosing the best fit based on correlation index will work well to discriminate the two phases in most cases. After comparison of the two files of fits to α- and β-Ti, 10817 pixels in the scan fitted better to α and 8283 to β. The scan area was cropped slightly before further analysis as some very thin material or possible vacuum was found in the bottom left corner of a scan (which gave very ambiguous and low quality results) by removing the leftmost and lowest 20 pixels giving a final analysed area with 6984 pixels of α and 6696 of β.

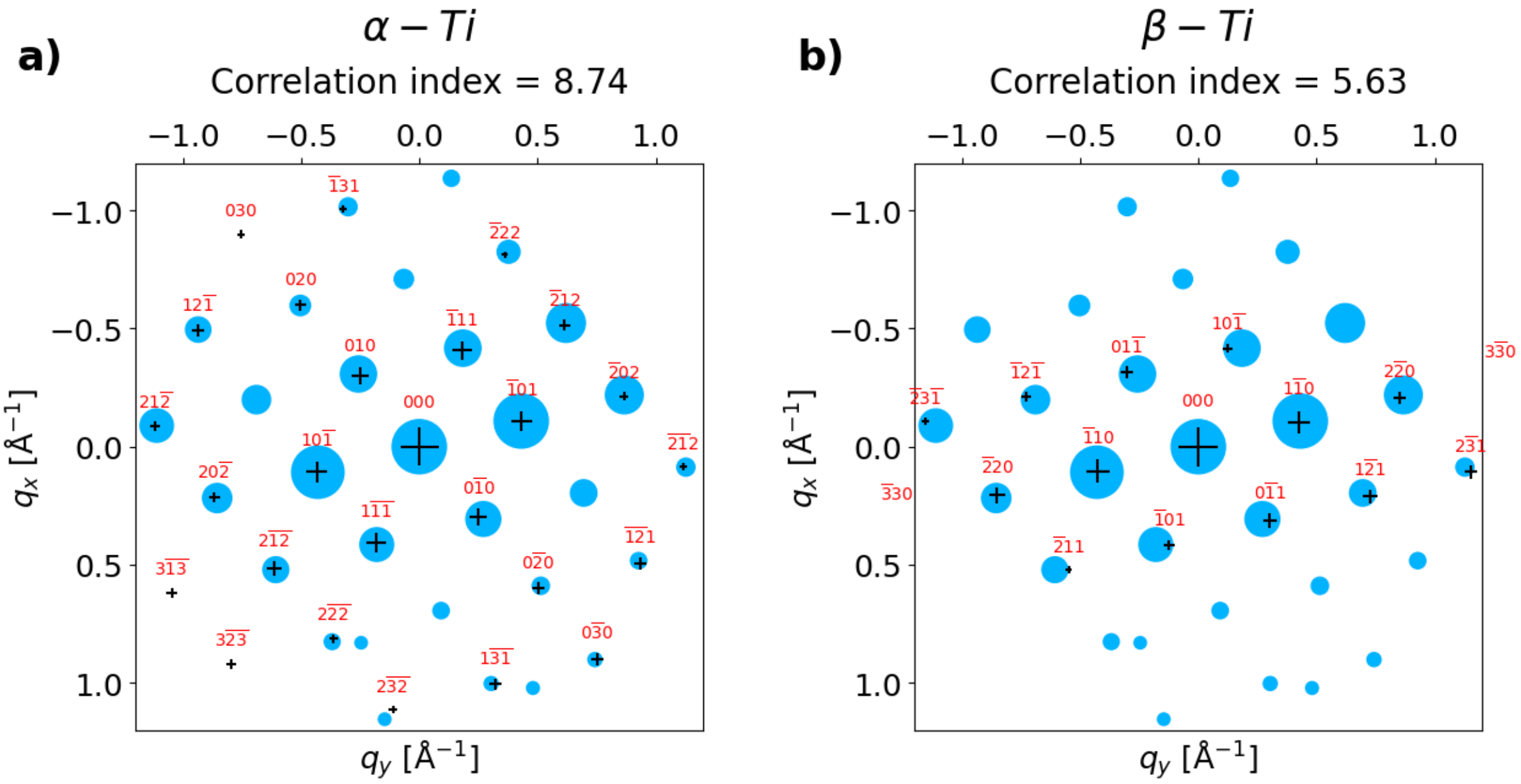

Figure 2: Indexing of the pattern (blue circles) shown in Figure 1f with the two phases: a) α-Ti; b) β-Ti (black crosses). The sizes of blue circles and black crosses is proportional to diffraction intensities. The correlation index is shown in both cases, and the fit is clearly rather better for the α-Ti. Note that the diffraction to scan rotation (of 34° counter clockwise) has been applied to align this with the axes of the images so the orientations of the patterns differ from 1f.

Figure 3 shows plots of crystallographic data calculated from the combined orientation dataset. Figures 3a-f are inverse pole figure (IPF) maps for the two phases. The α maps in Figure 3a-c show a number of different orientations but direct analysis of the exact orientations or their relationships to one another is not obvious from simply looking at the IPF maps. Nevertheless, it is clear that fine laths are well detected, down to areas 3 pixels wide, and thus less than 10 nm across. The β maps in Figure 3d-f all show a consistent orientation in this phase, which is perhaps unsurprising, suggesting that this was all a single $\beta$ crystal prior to the precipitation of the $\alpha$ laths (additionally, if the $\beta$ map alone is examined, i.e. indexing everything as if it were $\beta$ [shown in Figure S1 in the supplemental information], then the $\beta$ orientation appears consistent across the scan area, also suggesting that all the $\alpha$-laths originated from a single beta grain). To make the relationships between the $\alpha$-laths and the $\beta$ matrix clearer, there are two useful tools that can be used. Firstly, cluster analysis[22] was used on the dataset of all $\alpha$ orientations to determine clusters of similar orientations by cross-comparing all relative orientations, after applications of symmetry rotations to reduce them into the symmetry reduced zone for HCP. This produced 8 clusters containing 6915 orientations in total (2138, 730, 732, 610, 861, 1638, 153, and 53 in each cluster from 1-8), as well as a very small number of 69 outliers (not plotted). The average misorientation for each cluster could then be calculated, based on a misorientation from a given reference (one of the clusters). And each cluster was then assigned a colour and this colour plotted back into the map, which is shown in Figure 3g, which now makes it straightforward to see which laths have the same crystal orientation. Secondly, plotting the orientation data to pole figures is very helpful and this is done in Figure 3h and 3i. This shows that the $[0001]_{\alpha}$ poles of the laths are all aligned along one of the $\langle 110 \rangle_{\beta}$ poles, exactly as you would expect from the Burgers transformation ($[11\bar{2}0]_{\alpha} \parallel \langle 111 \rangle_{\beta}, (0001)_{\alpha} \parallel \langle 110 \rangle_{\beta}$. Moreover, many $\langle 2\overline{11}0 \rangle_{\alpha}$ directions are shown to align close to $\langle 111 \rangle_{\beta}$ directions, or be a few degrees off in Figure 3i (much as expected, e.g. in Figure 4 in Wang *et al.* [4]). Fairly obviously, not all 12 possible variants of the Burgers orientation relationship occur in this area. It should be noted that clusters 1 and 2, 3 and 4, and 5 and 6 all seem to occur in the same laths and seem to be related to each other. 3 and 4 have **c**-axes on opposite sides of the same stereogram with **a**-axes in the same places, and this just suggests a small tilt of the same orientation, possibly resulting from sample bending. 6 only occurs in one lath at the bottom and is mostly supplanted by 5, and all in laths with similar orientation. The points on the pole figures coincide and again this seems like slightly tilted versions of the same orientation. The only laths where there seems some confusion in orientation is those labelled 1/2 (gold/orange) where both orientations are apparently present. The pole figure shows that these have two $[0001]_{\alpha}$ directions symmetric about the centre, suggesting an ambiguity of how to assess where this **c**-direction points. Looking at Figure 1f reveals why this is, since the pattern is fairly close to having 2-fold symmetry (along a direction that does not have true 2-fold symmetry, just in the ZOLZ[29]) which leads to this ambiguity. As such, it is clear in hindsight that mapping orientations close to a high symmetry direction was a mistake and tilting off-axis a little would have broken this symmetry and may have led to more unambiguous results. Nevertheless, we can thus identify what appear to be 5 distinct lath orientations.

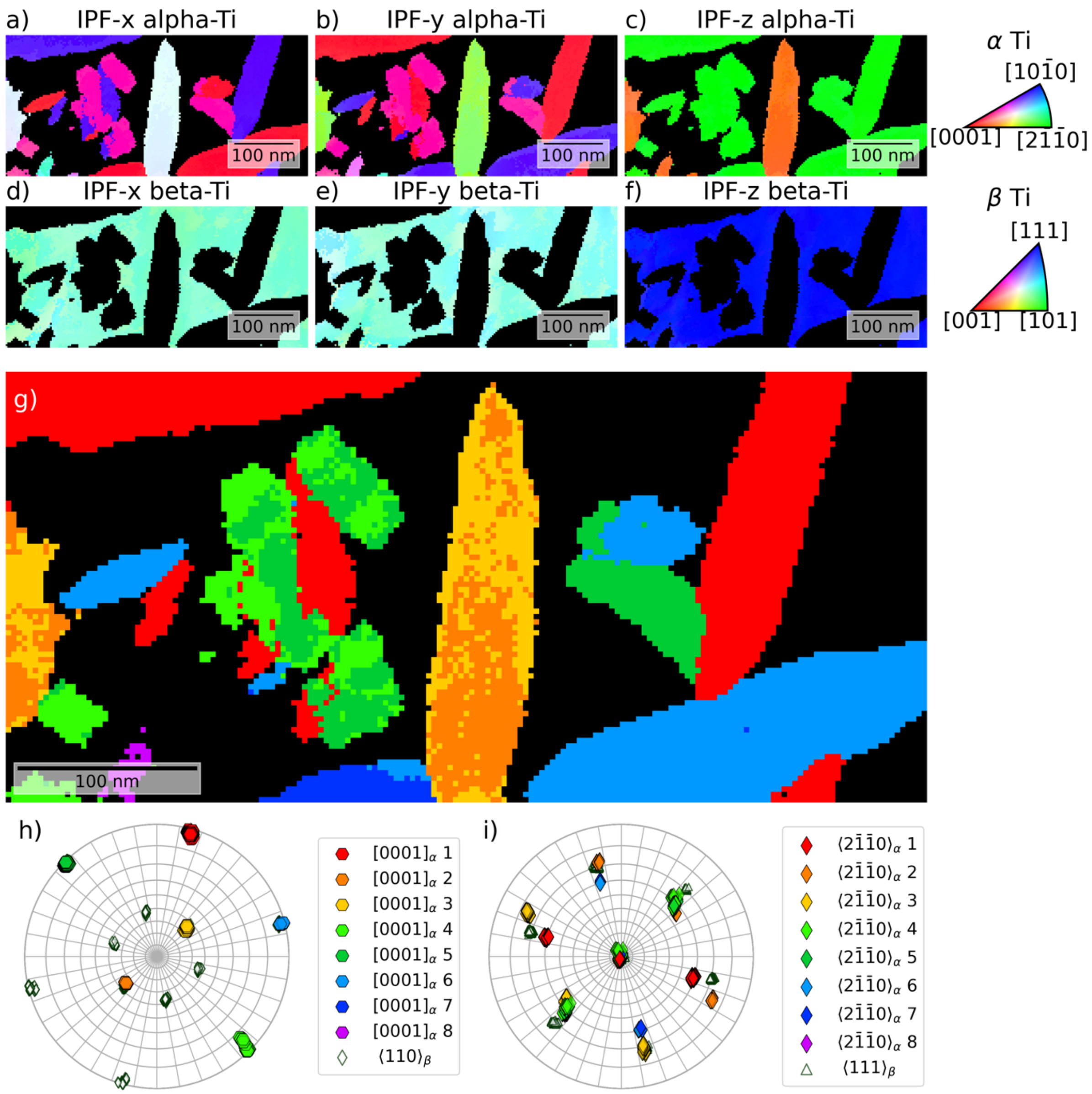

Figure 3: Mapping of α-Ti and β-Ti orientations in the scanned area as shown in Figure 1c): a)-c) α-Ti inverse pole figures for the x, y and z directions, with the colour key at the side; d)-f) β-Ti inverse pole figures for the x, y and z directions, with the colour key at the side; g) map of the distribution of the 8 distinct identified orientation clusters in the same colours as the pole figures; h) pole figure overlaying the $[0001]_\alpha$ directions (coloured hexagons) and $\langle 110\rangle_\beta$ directions (empty diamonds) with a colour key at the side i); pole figure overlaying the $\langle 2\bar{1}\bar{1}0\rangle_\alpha$ directions (coloured diamonds) and $\langle 111\rangle_\beta$ directions (empty triangles) with a colour key at the side.

Nevertheless, there is yet more information in the data, if one considers the relative orientations between the different variants, which are listed in Table 1 (calculated using average orientation for each cluster). These have been rationalised to one of the indices to match the calculations of the possible relative orientations between different alpha laths resulting from the Burgers transformation, as presented by Wang *et al.*[4]. These show that all the different clusters are related by relative orientations very close to these ideal orientation relationships.

| Orientation | u | v | w | t | θ (°) | 3σ (°) |
|---|---|---|---|---|---|---|
| 1 | 0 | 0 | 0 | 0 | 0 | 1.5 |
| 2 | -4.952 | -5.048 | 10.000 | -2.524 | 66.3 | 3.7 |
| 3 | -4.830 | 10.000 | -5.170 | 2.647 | 66.6 | 4.2 |
| 4 | -2.016 | 1.000 | 1.016 | -0.080 | 60.3 | 3.5 |
| 5 | -2.386 | 1.000 | 1.386 | -0.101 | 60.8 | 1.4 |
| 6 | -2.371 | 1.371 | 1.000 | -0.069 | 59.3 | 1.6 |
| 7 | -2.152 | 1.152 | 1.000 | -0.025 | 58.9 | 1.7 |
| 8 | 4.878 | 10.000 | -5.122 | -3.320 | 64.4 | 1.7 |

Table 1: Axis angle pairs ($[uvtw]/\theta$) for the average orientations of each orientation cluster characterising their misorientation from orientation 1. All have been normalised to the smallest *u, v,* or *t* index for comparison with the calculations of [4].

Orientations 1, 4/5 and 6/7 all have **c**-axes in the plane of the image, and roughly at 60° to each other and are related by roughly 60° rotations about something close to a $\langle 2\overline{11}0\rangle_\alpha$ direction (the one pointing vertical in this dataset, centre of the pole figure in Figure 3i) (type 2 orientation relationship of [4]). Moreover, the fact that there is a recurrence of three variants all related by rotations of 60° about one common **a**-direction in a so-called "triangle" configuration is something previously noted with coarser laths mapped with EBSD and well explained by [4] by consideration of shape strains for different combinations of laths.

Orientations 2 and 3, as said above, are the results of an ambiguous indexing and one of these two must be correct, but it is hard to be certain which from the data. Nevertheless, either gives a rotation from the orientation 1 (red in Figure 1g) of something close to the predicted 63.26° about a $\langle 10\overline{5}\overline{5}3\rangle_\alpha$ direction (type 4 of [4]). There seems no ambiguity about 8, which is just a single, small lath in the lower left, but that, too, corresponds to something close to this misorientation. This should be a frequently observed lath misorientation according to that work.

For reference, the 3σ value for the deviation within each cluster is also included in Table 1 to show the internal variation in each cluster. Although the initial cluster scan was set with what seems like a fairly broad criterion of members must be within 15° of each other, in actual fact, the resulting clusters are much tighter than this with most components within 1.5 to at most 4.2° (the 3σ criterion is that 99.7% of measurements will lie within this bound). Whilst this is a significantly larger misorientation range than the accuracy of ACOM itself, which should be ~1°, and probably better with precession[21], this allows for a degree of sample bending (as is normal in TEM) or local deviations from any theoretical relationship because of internal stresses and deformation. It should be noted that the 15° criterion is not absolute and needs to be experimented with for each dataset – choosing a larger number in this case resulted in fewer clusters and overlapped clusters of orientations that should be distinct, whereas choosing a smaller number just split some of these clusters into sub-clusters of very similar orientation (e.g. showing the same lamella as more than one cluster different parts of the scan because of the orientation changing with position as a result of sample bending).

Overall, this demonstrates that this is a useful workflow for performing automated crystallographic orientation mapping (ACOM) using scanning precession electron diffraction data in a complex two-phase microstructure and for performing quantitative crystallographic analysis on the results. Importantly, this goes beyond the single phase mapping of publications such as Cautaerts *et al.*[19] and Ophus *et al.*[21] or the single phase analysis with orix and clustering of Johnstone *et al.*[22]. Moreover, this works well at a mapping step size of 3 nm and reveals fine laths below 10 nm wide at their thinnest points, far beyond the spatial resolution limitations of EBSD. As such, this provides a valuable complement to EBSD (which will continue to be important for large sample areas) for studying finer laths and other nano-resolved crystallography. The fact that the orientation relationships here are expected to conform to a well-known one helps in assessing whether the ACOM is working correctly and producing plausible results. It would appear that going too close to a major zone axis in either phase could lead to 180° ambiguities in indexing. Unfortunately, whilst avoiding zone axes would be best, this cannot always be guaranteed, especially in larger scan areas of complex microstructures. Another thing that can be done is to reduce camera length, as higher angle diffraction spots not in the zero order Laue zone will not have the same issue of projection symmetry being higher than the true 3D space group symmetry[30]. However, this will not necessarily be so easy for small unit cells crystals like $\alpha$-Ti and $\beta$-Ti, as the higher order Laue zones will be at relatively high angle[24, 31] and this will be better at breaking ambiguity for crystals with larger unit cells. This will require that there is sufficient signal to noise ratio at the higher angles to still detect spots reliably. In that respect, the use of a direct electron counting detector, as in this work, is important, since we showed previously that the noise floor in older CCDs tends to obscure higher angle information [18]. In terms of the method, it would be possible to include comparison of the correlation indices and the writing of a single *.ang* file from *py4DSTEM* in the future. It may even be possible to consider cases where overlapped patterns are found as a linear combination of the two and then showing the proportions in each pixel, but those are innovations that can be explored in the future.

In the immediate future, this method will now be applied to studying the variant selection in the processing of two-phase titanium alloys and it is anticipated that more results will be published in due course, and in microstructures rather more complex that the one presented in this publication, although all the data acquisition, indexing, analysis and presentation methods developed here will be equally useful in that more complex situation.

**Conclusions**

Automated crystal orientation mapping of a two phase titanium alloy has been performed with recently introduced python libraries building on earlier work demonstrating their functionality in single phase materials. The alloy micro/nano-structure consisted of veins of $\beta$-Ti between large primary $\alpha$-Ti grains, these veins then containing smaller laths of secondary or tertiary $\alpha$-Ti. Highly reliable indexing was achieved of both phases (excepting a few minor ambiguities) allowing the construction of good quality orientation maps for both phases, allowing the mapping of various $\alpha$-Ti laths that have formed within a single beta grain. The orientations could be clustered into groups of similar orientations using automated cluster analysis, allowing the classification into just eight orientation clusters. Further analysis showed that three pairs of clusters related to similar orientations meaning that just five distinct crystal orientations for laths were present in the area analysed. Further analysis showed that all these orientations approximately follow the Burgers orientation relationship to the $\beta$ and show the expected relative orientations to one other. Three of the clusters follow expectations for a triangle of orientations related by ~60° rotations about a common $[11\bar{2}0]_\alpha$ axis, as has frequently been observed previously; which would correspond to a group of orientations with a low total shape strain. This work demonstrates a good analysis pathway for automated crystal orientation mapping of two- or multi-phase materials using transparent, Open-Source tools, and this will be applicable for more complex microstructures in titanium or other two- and multi-phase alloys.

**Acknowledgements**

We wish to acknowledge the use of the EPSRC funded Physical Sciences Data-science Service hosted by the University of Southampton and STFC under grant number EP/S020357/1 to allow access to the ICSD and the download of CIF files for appropriate structures. EFM is grateful to the EPSRC and TIMET for an industrial PhD studentship (EPSRC funding under EP/R512266/1). Helpful discussions and code upgrades with *orix* from Dr Patrick Harrison (SIMaP laboratory, Grenoble, France) and Mr Håkon Wiik Ånes (NTNU, Norway) are gratefully acknowledged. Work at the Molecular Foundry was supported by the Office of Science, Office of Basic Energy Sciences, of the U.S. Department of Energy under Contract No. DE-AC02-05CH11231.

**Data Deposit**

The raw data for the maps presented herein together with the Jupyter notebooks used for the processing are available for download at https://doi.org/10.5525/gla.researchdata.1514.